\documentclass[12pt, draftclsnofoot, onecolumn]{IEEEtran}
\usepackage{graphicx}
\usepackage{amsmath,bbm,amssymb,amsfonts,amstext,amsopn,sansmath}
\usepackage{cite}
\usepackage{balance}
\usepackage{url}
\usepackage{epsfig}
\usepackage{setspace}
\usepackage{stmaryrd}
\usepackage{psfrag}	
\usepackage{multirow}
\usepackage{float}
\usepackage[process=auto]{pstool}
\usepackage{etoolbox}
\usepackage{algorithm}
\usepackage{algorithmic}
\allowdisplaybreaks
\usepackage[nolist]{acronym}

\newcommand{\x}{\mathsf{x}}
\newcommand{\y}{\mathsf{y}}

\newcommand{\jj}{\mathsf{j}}
\newcommand{\Herm}{\mathsf{H}}
\newcommand{\e}{\mathsf{e}}

\def\by{\mathbf{y}}

\def\bH{\mathbf{H}}

\def\bA{\mathbf{A}}

\def\bx{\mathbf{x}}
\def\bn{\mathbf{n}}
\def\bG{\mathbf{G}}
\def\bD{\mathbf{D}}
\def\ba{\mathbf{a}}
\def\bd{\mathbf{d}}

\def\bPsi{\boldsymbol{\Psi}}
\def\bOmega{\boldsymbol{\Omega}}
\def\bSigma{\boldsymbol{\Sigma}}

\def\Cset{\mathbb{C}}

\newcommand{\diag}{\mathrm{diag}}

\newtheorem{lem}{Lemma}

\newtheorem{prop}{Proposition}
\newtheorem{corol}{Corollary}

\newtoggle{OneColumn}
\toggletrue{OneColumn}

\newtoggle{arXiv}
\togglefalse{arXiv}

\EndPreamble
\begin{document}
\title{Power Efficiency, Overhead, and Complexity Tradeoff in IRS-Assisted Communications -- \\ Quadratic Phase-Shift Design}

\author{Vahid Jamali, \textit{Member, IEEE}, Marzieh Najafi, \textit{Student Member, IEEE}, \\  Robert Schober, \textit{Fellow, IEEE}, and H. Vincent Poor, \textit{Life Fellow, IEEE} \vspace{-0.7cm}%
	\thanks{V. Jamali, M. Najafi, and R. Schober are with the Institute for Digital Communications at Friedrich-Alexander University Erlangen-N\"urnberg (FAU) (e-mail:
	vahid.jamali@fau.de;  marzieh.najafi@fau.de;		robert.schober@fau.de).}
\thanks{H. Vincent Poor is with the Department of Electrical Engineering, Princeton University, Princeton, NJ 08544 USA (e-mail: poor@princeton.edu).}
}

\maketitle

\begin{abstract}
In this paper, we focus on large intelligent reflecting surfaces (IRSs) and propose a new codebook construction method to obtain a set of predesigned phase-shift configurations for the IRS unit cells. Since the overhead for channel estimation and the complexity of online optimization for IRS-assisted communications scale with the size of the phase-shift codebook, the design of small codebooks is of high importance. We show that there exists a fundamental tradeoff between  power efficiency and the size of the codebook. We first analyze this tradeoff for  baseline designs that employ a linear phase-shift across the IRS. Subsequently,  we show that an efficient design for small codebooks mandates higher-order phase-shift variations across the IRS. Consequently, we propose a quadratic phase-shift~design, derive its coefficients as a function of the codebook size,~and~analyze~its performance. Our simulation results show that the proposed design yields a higher  power  efficiency for small codebooks than the linear baseline~designs. 
\end{abstract}
\vspace{-0.3cm}

\acresetall
\section{Introduction}

Intelligent reflecting surface- (IRS-) assisted communication is a promising emerging technology for the realization of smart wireless environments \cite{di2019smart}. IRSs consist of a large number of programmable sub-wavelength elements, so-called unit cells or meta atoms, that can change the properties of an impinging electromagnetic wave while reflecting it \cite{jamali2019intelligent}. For instance, a properly designed unit-cell phase distribution across the
surface allows the IRS to alter the direction of the wavefront of the reflected wave, thereby realizing the generalized Snell's law  \cite{estakhri2016wave,najafi2020intelligentoptic}. However, it has been shown in the literature \cite{bjornson2020power,di2020analytical,najafi2020intelligent} that,  for far-field applications, IRS-assisted links are effective only if the IRS is very large (e.g., comprising hundreds if not thousands of unit cells). Such large IRSs introduce various implementation challenges including large overhead for channel estimation  and high complexity for online optimization.

To cope with the aforementioned challenges, the authors  in \cite{najafi2020intelligent} proposed to not directly optimize the unit-cell phase shifts, but to design  a set of  phase-shift configurations, referred to as transmission modes,  in an offline stage, and select the best transmission mode in the online stage for a given channel realization. Similarly, in \cite{zheng2019intelligent,you2020channel,alwazani2020intelligent}, the authors employed  predefined phase-shift configurations based on the discrete Fourier transform (DFT) matrix for channel estimation in IRS-assisted systems. We refer to a model that characterizes the IRS in terms of  the phase-shift applied by its individual unit cells as the \textit{phase-shift model} and to a model that characterizes the IRS in terms of transmission modes  as the \textit{transmission-mode model}. In contrast to the phase-shift model, whose channel estimation overhead and online optimization complexity scale with the number of IRS unit cells, in the transmission-mode model, these quantities scale with the number of transmission modes contained in the codebook which, in principle, can be much smaller than the number of IRS unit cells. Nevertheless, the codebooks proposed in \cite{najafi2020intelligent,zheng2019intelligent,you2020channel,alwazani2020intelligent} may still be too big for practical implementation (e.g., the codebooks in \cite{najafi2020intelligent} and \cite{zheng2019intelligent,you2020channel,alwazani2020intelligent} contain $100$ and $400$ transmission modes, respectively, for an IRS of size $10\lambda\times10\lambda$, where $\lambda$ denotes the wavelength).

Motivated by the above discussion, in this paper, we aim to design small-size but efficient phase-shift codebooks. To this end, we show that there is a fundamental tradeoff between the codebook size and the resulting transmission power efficiency  such that the smaller the codebook is, the lower the power efficiency of the IRS-assisted system becomes. We first analyze this tradeoff for the baseline schemes in \cite{najafi2020intelligent,zheng2019intelligent,you2020channel,alwazani2020intelligent}, where a \textit{linear phase shift} across the IRS was considered. Then, we show that for small phase-shift codebooks and large IRSs, higher-order variations of the phase shifts  across the IRS are needed. In particular, we propose a \textit{quadratic phase-shift design} where the phase shifts across the IRS are quadratic functions and derive its coefficients as a function of the desired codebook  size. Moreover, we derive a closed-form approximation of the IRS response function for the   proposed phase-shift design that characterizes the response of the IRS in the far-field. Furthermore, we study the tradeoff between power efficiency and codebook size for the proposed design as well as the baseline designs via computer simulations.

We note that the problem of phase-shift codebook design for IRS-assisted communication is similar to the problem of beam synthesis in millimeter-wave communications \cite{cao2016constant,palacios2016lightweight,de2017millimeter}. However,  unlike  millimeter-wave  systems where the signal is known to the transmitter and the problem is to synthesize a  transmission pattern along certain directions, in IRS-assisted communication,  the phase distribution of  the incident wave across the IRS is unknown when synthesizing the desired reflection  pattern. As a result, the beam designs available in the millimeter-wave communication literature  are not directly  applicable to IRS phase-shift codebook design.


\section{System Model}

In this paper, we focus on the problem of codebook design based on the transmission-mode model in \cite{najafi2020intelligent}. Nevertheless, for completeness, we present both the phase-shift and transmission-mode models for an example IRS-assisted downlink system. This  makes the relation between this work and the literature, which is mostly based on the phase-shift model,~clearer.
 
\subsection{Phase-shift Model}
Let us consider an example downlink network consisting of a base station (BS) equipped with $N_t$ transmit antennas serving $K$ mobile users each equipped with $N_r$ receive antennas. The direct links between the BS and the users may exist but are severely shadowed (e.g., by a blocking building). The downlink communication is assisted by an IRS that comprises $Q$  reflecting unit cells. The end-to-end system model can be written as \cite{najafi2020intelligent}  
\begin{IEEEeqnarray}{ll}\label{Eq:model_phase}
\by_k & = (\bH_{d,k} + \bH_{r,k} \bOmega\bH_i)\bx  + \bn_k ,\quad k = 1,\ldots,K,
\end{IEEEeqnarray}
where $\bx\in\Cset^{N_t}$, $\by_k\in\Cset^{N_r}$, and $\bn_k\in\Cset^{N_r}$
denote the BS's transmit signal, the received signal at the $k$-th user, and the additive white Gaussian noise (AWGN) 
at the $k$-th user, respectively. Here, $\Cset$ denotes the set of complex numbers. Moreover, $\bH_{d,k}\in\Cset^{N_r\times N_t}$, $\bH_i\in\Cset^{Q\times N_t}$, and $\bH_{r,k}\in\Cset^{N_r\times Q}$ denote the BS-to-user~$k$, BS-to-IRS, and IRS-to-user~$k$ channel matrices, respectively.  Furthermore, $\bOmega=\diag(\bar{g}\e^{\jj \omega_1},\dots,\bar{g}\e^{\jj \omega_Q})\in\Cset^{Q\times Q}$ is a diagonal matrix with diagonal entries $\bar{g}\e^{\jj \omega_1},\dots,\bar{g}\e^{\jj \omega_Q}$, where $\omega_q$ is the phase-shift applied by the $q$-th unit cell and $\bar{g}$ denotes the  normalized unit-cell factor.  For simplicity, we assume a constant unit-cell factor  $\bar{g}=\frac{4\pi A_{\rm uc}}{\lambda^2}$, where $A_{\rm uc}$  is  the unit-cell area; however, in general, $\bar{g}$ may also depend on the  incident and reflected angles of the waves, see \cite{di2020analytical,najafi2020intelligent} for details. 

Since $Q$ is typically very large, the elements of $\bH_{r,k}$ (and $\bH_i$) cannot be modeled as independent random variables and the rank of $\bH_{r,k}$ (and $\bH_i$) is limited by the number of channel scatters. Adopting the low-rank channel model in \cite{najafi2020intelligent}, we decompose the channel matrices as $\bH_i = \bA_{i} \bSigma_{i} \bD_{i}^\Herm$, $\bH_{r,k} = \bA_{r,k} \bSigma_{r,k} \bD_{r,k}^\Herm$, and $\bH_{d,k} = \bA_{d,k} \mathbf{\Sigma}_{d,k} \bD_{d,k}^\Herm$, respectively, where the components $\{\bA_{i} \in \Cset^{Q \times {L_i}} $, $\bA_{r,k} \in \Cset^{N_r \times {L_{r,k}}}$, $\bA_{d,k} \in \Cset^{N_r \times {L_{d,k}}}\} $, $\{\mathbf{D}_{i} \in \Cset^{N_t \times {L_i}}$, $\bD_{r,k} \in \Cset^{Q \times {L_{r,k}}}$, $\bD_{d,k}\in \Cset^{N_{t} \times {L_{d,k}}}\}$, and $\{\bSigma_{i}\in \Cset^{L_i \times {L_i}}$, $\bSigma_{r,k} \in \Cset^{L_{r,k} \times {L_{r,k}}}$, $\bSigma_{d,k}\in \Cset^{L_{d,k} \times {L_{d,k}}}\} $ represent the receive steering matrices (corresponding to the angles-of-arrival (AoAs)), transmit steering matrices (corresponding to the angles-of-departure (AoDs)), and channel gains of the scatters, where  $L_i, L_{r,k},$ and  $L_{d,k}$ denote the numbers of scatters of the BS-to-IRS,  IRS-to-user~$k$, and BS-to-user~$k$ channels, respectively. Furthermore, $(\cdot)^\Herm$ denotes the Hermitian transpose. 

\subsection{Transmission-mode Model}


The transmission-mode model introduced in \cite{najafi2020intelligent} relies on  designing a predefined set of $M$ phase-shift configurations in an offline stage which later can be selected  for  online transmission or channel estimation\footnote{To facilitate flexible optimization of the IRS, in \cite{najafi2020intelligent}, it was proposed to divide the IRS unit cells into several groups, called tiles, and to devise transmission modes for each tile. In this case, phase-shift codebooks designed in this paper can be applied for each tile.}. Let $g_{m}(\bPsi_i,\bPsi_r)$ denote the  normalized IRS response function for the $m$-th transmission mode, where $\bPsi_i=(\theta_i,\phi_i)$ and  $\bPsi_r=(\theta_i,\phi_r)$ are the AoA of the incident wave and the AoD of the reflected wave, respectively, and $\theta$ and $\phi$ are used to denote the elevation and azimuth angles, respectively. The end-to-end system model in terms of the IRS response function can be rewritten for $k = 1,\ldots,K$ as follows~\cite{najafi2020intelligent} 
\begin{IEEEeqnarray}{ll}\label{Eq:model_mode}
\by_k =  \Big(\bA_{d,k} \mathbf{\Sigma}_{d,k} \bD_{d,k}^\Herm + \bA_{r,k} \bSigma_{r,k}  \bG_k  \bSigma_{i} \bD_{i}^\Herm 
 \Big)\bx  + \mathbf{n}_k ,
\end{IEEEeqnarray}
where $ \bG_k \in \Cset^{L_{r,k} \times L_{i}}$ is given by 
\begin{IEEEeqnarray}{ll}\label{Eq:response}
\bG_k = \bD_{r,k}^\Herm\bOmega\bA_{i} 
= \sum_{m=1}^M s_{m} \bG_{m,k}.\quad 
\end{IEEEeqnarray}
Here, the element in the $n_r$-th row and $n_i$-th column of $\bG_{m,k}=\bD_{r,k}^\Herm\bOmega_{m}\bA_{i}$ is the IRS response function $g_{m}(\bPsi_i,\bPsi_r)$ evaluated at the $n_i$-th AoA of the IRS and $n_r$-th AoD from the IRS to the $k$-th user.   Moreover, $\bOmega_{m}$ is the phase-shift matrix for the $m$-th transmission mode. Furthermore, $s_{m}\in\{0,1\}$ is a binary variable which is equal to one if the $m$-th mode is selected and otherwise it is equal to zero. Since only one mode can be selected, $\sum_{m=1}^Ms_{m}=1$ has to hold. Throughout the paper, we assume that the IRS is a planar uniform array with $Q_\x$ ($Q_\y$) unit cells spaced $d_\x$ ($d_\y$)  a apart on the $x$-axis ($y$-axis), indexed by  $n_\x=0,\dots,Q_\x-1$ ($n_\y=0,\dots,Q_\y-1$), where  $Q=Q_\x Q_\y$ and each unit cell has an area of $A_{\rm uc}=d_\x d_\y$. Moreover, for future reference, we define $A_\x(\bPsi_i,\bPsi_r)=A_\x(\bPsi_i)+A_\x(\bPsi_r)$ and $A_\y(\bPsi_i,\bPsi_r)=A_\y(\bPsi_i)+A_\y(\bPsi_r)$ with $A_\x(\bPsi)=\sin(\theta)\cos(\phi)$ and $A_\y(\bPsi)=\sin(\theta)\sin(\phi)$.

\subsection{Power Efficiency, Overhead, and Complexity}

In the following, we discuss the fundamental tradeoff between power efficiency, channel estimation overhead, and complexity of online IRS optimization  in terms of the size of the codebook for the unit-cell phase shifts. 

\textbf{Power efficiency:} In order to maximize the power reflected in the direction where the user is located, the IRS has to introduce an appropriate phase shift at each unit cell such that the waves propagating from all unit cells in the direction of interest  add up coherently. This yields the maximum value of $|g_{m}(\bPsi_i,\bPsi_r)|$ along the user direction, which, in \cite[Corollary~2]{najafi2020intelligent}, is shown to be equal to $g^{\max}=\bar{g}Q_\x Q_\y$. 
The corresponding unit-cell phase shifts depend on the AoA of the incident wave and the desired AoD in which the user is located, which are  both real numbers. In order to construct a finite-size codebook,  the AoA and AoD have to be discretized. Therefore, for given AoA and AoD, it is not guaranteed that there exists a transmission mode in the codebook  whose value of the IRS response function is equal to $g^{\max}$. We characterize this reduction of received power due to the finite-size codebook by a \textit{power efficiency} factor which is formally defined~as 
\begin{IEEEeqnarray}{ll}\label{Eq:pEE}
	\gamma(\bPsi_i,\bPsi_r) 
	= \underset{m=1,\dots,M }{\max}\left[\frac{|g_{m}(\bPsi_i,\bPsi_r) |}{\bar{g}Q_\x Q_\y}\right]^2.
\end{IEEEeqnarray}
Note that by definition,  $0\leq\gamma(\bPsi_i,\bPsi_r) \leq 1$~holds.

\textbf{Channel estimation overhead:} A common approach for codebook-based channel estimation is to have the IRS select a  given phase-shift configuration from the codebook,  the transmitter  send pilot symbols, and the receiver estimate the end-to-end channel  \cite{zheng2019intelligent,you2020channel,alwazani2020intelligent}. This procedure is repeated until the channel is estimated for all phase-shift configurations in the codebook. Therefore, the channel estimation overhead is directly proportional to the size of the phase-shift codebook $M$. 

\textbf{Complexity of online optimization:} Based on the estimated channel, the best transmission mode among those in the codebook can be selected for data transmission during online optimization.  In principle, the complexity of online optimization scales linearly with the size of the codebook $M$; however, efficient algorithms can be developed to further reduce the complexity, see e.g. the mode pre-selection  algorithm in \cite{najafi2020intelligent}.

In summary, the larger the codebook size, the higher the power efficiency but also the larger the channel estimation overhead and the complexity of online resource allocation.

\section{Phase-shift Design}

In this section, we focus on the channel model described in \eqref{Eq:model_mode}, introduce different phase-shift designs, and analyze the resulting IRS response functions as well as the tradeoff between power efficiency and  codebook~size.

\subsection{Baseline Schemes}

\subsubsection{DFT-based Design}
A common choice for the IRS unit-cell phase shifts are the columns of the DFT matrix \cite{zheng2019intelligent,you2020channel,alwazani2020intelligent}. Assuming a unit-cell spacing of $d_\x=d_\y=\frac{\lambda}{2}$, the phase shift applied by the $(n_\x,n_\y)$-th unit cell, denoted by $\omega_{n_\x,n_\y}$, is given~by
\begin{IEEEeqnarray}{ll}\label{Eq:DFT}
	\e^{\jj\omega_{n_\x,n_\y}}
	= \e^{-\frac{\jj2\pi }{Q_\x}m_\x n_\x} 
	\times \e^{-\frac{\jj2\pi}{Q_\y}m_\y n_\y},
\end{IEEEeqnarray}
where $m_t=0,\dots,Q_t-1,\,\,t\in\{\x,\y\}$. Here, each pair $(m_\x,m_\y)$ constitutes one element of the phase-shift codebook, i.e., the codebook size is $M=Q=Q_\x Q_\y $. As can be observed, \eqref{Eq:DFT} represents the elements of a two-dimensional DFT matrix. From \eqref{Eq:response}, the IRS response function is obtained as
\begin{IEEEeqnarray}{ll}\label{Eq:gfunc}
	g_{m}(\bPsi_i,\bPsi_r)  
	&= \bd_{r}^\Herm\bOmega_{m}\ba_{i} 
	\nonumber\\
	&=\bar{g}\sum_{n_\x=0}^{Q_\x-1}\sum_{n_\y=0}^{Q_\y-1}
	\e^{\frac{\jj 2\pi d_\x A_\x(\bPsi_i,\bPsi_r)}{\lambda}n_\x}
	\e^{\frac{\jj 2\pi d_\y A_\y(\bPsi_i,\bPsi_r)}{\lambda}n_\y}
	\e^{\jj\omega_{n_\x,n_\y}},\quad
\end{IEEEeqnarray}
 where $\ba_{i}$ and $\bd_{r}$ are the steering vectors for AoA $\bPsi_i$ and AoD $\bPsi_r$, whose elements corresponding to the $(n_\x,n_\y)$-th unit cell are given by $\e^{\frac{\jj 2\pi (d_\x A_\x(\bPsi_i)n_\x+d_\y A_\y(\bPsi_i)n_\y)}{\lambda}}$ and $\e^{-\frac{\jj 2\pi (d_\x A_\x(\bPsi_r)n_\x+d_\y A_\y(\bPsi_r)n_\y)}{\lambda}}$, respectively. The following lemma gives the IRS response function  $g_{m}(\bPsi_i,\bPsi_r)$ for the $(m_\x,m_\y)$-element of the codebook, denoted by $g_{(m_\x,m_\y)}(\bPsi_i,\bPsi_r)$.

\begin{lem}\label{Lem:gLinear}
Given the DFT-based phase-shift design in \eqref{Eq:DFT} and a wave impinging on the IRS from AoA $\bPsi_i$, the  IRS has the following response function for AoD $\bPsi_r$
\begin{IEEEeqnarray}{ll}\label{Eq:gDFT}
	g_{(m_\x,m_\y)}(\bPsi_i,\bPsi_r) 
	= \bar{g} \e^{-\jj(Q_\x-1)\varpi_\x}\e^{-\jj(Q_\y-1)\varpi_\y} \times\frac{\sin(Q_\x\varpi_\x)}{\sin(\varpi_\x)}
	\times\frac{\sin(Q_\y\varpi_\y)}{\sin(\varpi_\y)},
\end{IEEEeqnarray}
where $\varpi_t=\frac{\pi d_t A_t(\bPsi_i,\bPsi_r)}{\lambda}-\frac{\pi m_t}{Q_t},\,\,t\in\{\x,\y\}$.
\end{lem}
\begin{IEEEproof}
	The proof follows the same steps as a similar proof in \cite[Appendix~B]{najafi2020intelligent} and is omitted due to space constraints.
\end{IEEEproof}

\begin{corol}\label{Corol:PeffDFT}
	The normalized power efficiency of  the DFT-based codebook design in \eqref{Eq:DFT} is bounded as 
	\begin{IEEEeqnarray}{ll}\label{Eq:gDFTpower}
		\frac{16}{\pi^4}
		\overset{(a)}{\leq}
		\Bigg[\frac{1}{Q_\x Q_\y\sin\left(\frac{\pi}{2Q_\x}\right)\sin\left(\frac{\pi}{2Q_\y}\right)}\Bigg]^2\!\!
		\leq\gamma(\bPsi_i,\bPsi_r)  \overset{(b)}{\leq} 1,\quad
	\end{IEEEeqnarray}
	where inequality $(a)$ holds with equality as $Q_\x,Q_\y\to\infty$ and inequality $(b)$ holds with equality for AoAs and AoDs for which $A_t(\bPsi_i,\bPsi_r) = \frac{m_t \lambda}{d_tQ_t},\,\,t\in\{\x,\y\}$, holds.
	\end{corol}
\begin{IEEEproof}
	The maximum value of $\gamma(\bPsi_i,\bPsi_r)$ is $1$ and occurs at $\varpi_t=0,\,\,t\in\{\x,\y\}$. Noticing the monotonically decreasing property of function $\frac{\sin(Ax)}{\sin(x)}$ around $x=0$, the minimum value of  $\gamma(\bPsi_i,\bPsi_r)$ occurs when $\varpi_\x$ and $\varpi_\y$ attain their maximum values which are given by $\varpi_t=\frac{\pi}{2Q_t},\,\,t\in\{\x,\y\}$, whose substitution in \eqref{Eq:gDFT} leads to the tighter lower bound in \eqref{Eq:gDFTpower}. The lower bound in $(a)$ follows directly from $\sin(x)\leq x,\,\,x\geq0$.
\end{IEEEproof}

As can be seen from \eqref{Eq:gDFTpower}, the normalized power efficiency of the DFT-based codebook design cannot be smaller than $\frac{16}{\pi^4}\approx0.1643$ for any AoA and AoD in the asymptotic regime of large $Q_\x$~and~$Q_\y$.

\subsubsection{Linear Phase-shift Design \cite{najafi2020intelligent}} 
The phase shift $\omega_{n_\x,n_\y}$ in \eqref{Eq:DFT} is a linear function in variables $n_\x$ and $n_\y$ with constant slopes of $\frac{2\pi m_\x}{Q_\x}$  and $\frac{2\pi m_\y}{Q_\y}$, respectively. Therefore, the codebook size is always fixed as $M=Q_\x Q_\y$ and the power efficiency is given in \eqref{Eq:gDFTpower}. Moreover, \eqref{Eq:DFT}  is valid for a unit-cell spacing of a half wavelength. In \cite{najafi2020intelligent}, the authors considered a general linear phase shift for  IRSs with general unit-cell spacing and derived the corresponding IRS response function. Next, we restate the design in \cite{najafi2020intelligent} with a slight change of notation. In particular, the linear phase design in  \cite{najafi2020intelligent} can be rewritten as 
\begin{IEEEeqnarray}{ll}\label{Eq:Linear}
	\e^{\jj\omega_{n_\x,n_\y}}
	&= \e^{-\frac{\jj2\pi d_\x\bar{\beta}_\x}{M_\x\lambda}m_\x n_\x} 
	\times \e^{-\frac{\jj2\pi d_\y\bar{\beta}_\y}{M_\y\lambda}m_\y n_\y},
\end{IEEEeqnarray}
where $m_t=0,\dots,M_t-1,\,\,t\in\{\x,\y\}$, $M_\x M_\y=M$, and $\bar{\beta}_t=\min\{4,\lambda/d_t\},\,\,t\in\{\x,\y\}$. Here,  $\bar{\beta}_t$ specifies the range of the phase-shift gradients needed in the codebook for transforming any AoA into any AoD for a unit-cell spacing of $d_t,\,\,t\in\{\x,\y\}$.  Note that both \eqref{Eq:DFT} and \eqref{Eq:Linear} are linear phase shifts where \eqref{Eq:Linear}  reduces to \eqref{Eq:DFT} for $M_\x=Q_\x$, $M_\y=Q_\y$, and $d_\x=d_\y=\frac{\lambda}{2}$.
The IRS response function for the phase shift in  \eqref{Eq:Linear} is identical to that given in Lemma~\ref{Lem:gLinear} after replacing $\varpi_t$ with $\varpi_t=\frac{\pi d_t A_t(\bPsi_i,\bPsi_r)}{\lambda}-\frac{\pi d_t\bar{\beta}_t m_t}{M_t\lambda},\,\,t\in\{\x,\y\}$.
Note that the resulting IRS response function is consistent with that given in \cite[Proposition~2]{najafi2020intelligent}. Next, we analyze the power efficiency of the linear design in  \eqref{Eq:Linear} in terms of  codebook size~$M$. 

\begin{corol}\label{Corol:Pefflinear}
	The normalized power efficiency of  the linear phase-shift codebook design  is bounded as 
	\begin{IEEEeqnarray}{ll}\label{Eq:gLinearpower}
		\Bigg[\frac{\sin\left(\frac{\pi d_\x\bar{\beta}_\x Q_\x}{2M_\x\lambda}\right)\sin\left(\frac{\pi d_\x \bar{\beta}_\y Q_\y}{2M_\y\lambda}\right)}{Q_\x Q_\y\sin\left(\frac{\pi d_\x\bar{\beta}_\x}{2M_\x\lambda}\right)\sin\left(\frac{\pi d_\y\bar{\beta}_\y}{2M_\y\lambda}\right)}\Bigg]^2\!\!
		\overset{(a)}{\leq}\gamma(\bPsi_i,\bPsi_r)  \overset{(b)}{\leq} 1,\quad
	\end{IEEEeqnarray}
	 where inequality $(a)$ holds for $M_t\geq \frac{d_t\bar{\beta}_tQ_t}{2\lambda},\,\,t\in\{\x,\y\}$ and inequality $(b)$ holds with equality for AoAs and AoDs for which $A_t(\bPsi_i,\bPsi_r) = \frac{\bar{\beta}_t m_t}{M_t},\,\,t\in\{\x,\y\}$. Moreover, for $M_\x,M_\y\to\infty$, the power efficiency  approaches one for all AoAs and AoDs.  Furthermore, for $M_t\leq \frac{d_t\bar{\beta}_tQ_t}{2\lambda},\,\,t\in\{\x,\y\}$, there exist AoAs and AoDs which are not supported by the main lobe of $g_m(\bPsi_i,\bPsi_r)$ for any of the transmission modes in the codebook.
\end{corol}
\begin{IEEEproof}
	The proof is similar to the proof of  Corollary~\ref{Corol:PeffDFT} and is omitted due to space constraints.
\end{IEEEproof}

Corollary~\ref{Corol:Pefflinear} reveals the tradeoff between  power efficiency $\gamma(\bPsi_i,\bPsi_r)$ and codebook size $M$ for the linear phase-shift design in \eqref{Eq:Linear}. In particular, the normalized power efficiency may approach one by increasing   $M_\x,M_\y\to\infty$; however, this implies a large overhead for channel estimation and high complexity for online optimization. In fact, in practice, we are more interested in  small values of $M_\x$ and $M_\y$, where  Corollary~\ref{Corol:Pefflinear} states that assuming half-wavelength unit-cell spacing, $M_\x$ and $M_\y$ cannot be smaller than $\frac{Q_\x}{2}$ and $\frac{Q_\y}{2}$, respectively, otherwise, the power efficiency becomes very small for some AoA and AoD pairs  that are supported only by the side lobes of $g_m(\bPsi_i,\bPsi_r),\,\,\forall m$. In other words,  $M>\frac{Q_\x Q_\y}{4}$ has to hold which is particularly limiting for large IRSs.  To cope with this issue, in the following, we propose a novel phase-shift design whereby $M$ may assume values much smaller than $\frac{Q_\x Q_\y}{4}$,  while the power efficiency still remains large for all AoAs and AoDs.

\subsection{Proposed Quadratic Phase-shift Design} 

In the following, we first present the basic idea behind the proposed phase-shift design. Subsequently, we analyze its performance in terms of the IRS response function and the tradeoff between power efficiency and codebook size.

\subsubsection{Basic Idea Behind the Proposed Phase-shift Design}
Recall that a linear phase shift  transforms a plane wave with AoA $\bPsi_i$ into another wave propagating into the direction of a certain AoD $\bPsi_r$. However, for a finite-size codebook, we require that for each codebook element $m$, the IRS transforms the AoAs in a given \textit{interval} $\bPsi_i\in\mathcal{A}_{i,m}$ into the AoDs in  another  \textit{interval} $\bPsi_r\in\mathcal{A}_{r,m}$.  In fact, for any pair   $(\bPsi_i,\bPsi_r)\in\mathcal{A}_{i,m}\times\mathcal{A}_{r,m}$, the phase-shift function $\omega_{n_\x,n_\y}$  should have a component with the corresponding phase gradient for the transformation of a wave with AoA $\bPsi_i$ into a wave with AoD $\bPsi_r$. Obviously, this is not attainable with a \textit{linear} phase-shift design. Let us define  the normalized phase-shift gradient $\frac{\partial \omega_{n_\x,n_\y}}{\partial n_t} \triangleq \frac{2\pi d_t}{\lambda}\beta_{t}(n_t)$ and the corresponding interval  $\beta_{t}(n_t)\in\mathcal{B}_{t,m_t}\subset[0,\bar{\beta}_t),\,\,t\in\{\x,\y\}$, for the $(m_\x,m_\y)$-th transmission mode, where $\bar{\beta}_t=\min\{4,\lambda/d_t\},\,\,t\in\{\x,\y\}$.
 Since the normalized phase-shift gradient  varies across the  surface, i.e., $\beta_{t}(n_t)\in\mathcal{B}_{t,m_t}$, we require a higher-order variation of the phase shift across the IRS compared to the linear phase-shift design. Hereby, we propose the following \textit{quadratic} phase-shift function
\begin{IEEEeqnarray}{ll}\label{Eq:Quadratic}
	\e^{\jj\omega_{n_\x,n_\y}}
	\!	=\! \e^{\!-\frac{\jj 2\pi d_\x}{\lambda} \!\big[\frac{\Delta\beta_{\x,m_\x}}{2Q_\x}n_\x^2+\beta_{\x,m_\x} n_\x\big]} 
	\! \e^{\!-\frac{\jj 2\pi d_\y}{\lambda} \!\big[\frac{\Delta\beta_{\y,m_\y}}{2Q_\y }n_\y^2+\beta_{\y,m_\y} n_\y\big]}\!,\quad\,\,\,\,
\end{IEEEeqnarray}
which implies $\mathcal{B}_{t,m_t}=[\beta_{t,m_t},\beta_{t,m_t}+\Delta\beta_{t,m_t}],$ $\forall m_t,t\in\{\x,\y\}$.  Note that $\Delta\beta_{t,m_t}=0$ implies a linear phase-shift design, i.e., the baseline linear phase-shift designs can be seen as special cases of the proposed quadratic design. In this paper, we focus on a uniform quantization of the phase-shift gradient leading to $\beta_{t,m_t+1}=\beta_{t,m_t}+\Delta\beta_{t}$ and $\Delta\beta_{t,m_t}=\Delta\beta_{t}=\frac{\bar{\beta}_t}{M_t},\,\, m_t=0,\dots,M_t-1,t\in\{\x,\y\}$.

\subsubsection{IRS Response Function} 
Unfortunately, substituting the quadratic phase-shift in  \eqref{Eq:Quadratic} into \eqref{Eq:gfunc} does not yield a closed-form solution for the IRS response function. To cope with this issue, we approximate the summations in \eqref{Eq:gfunc} with the corresponding integrations which yields an approximation  of $g_{(m_\x,m_\y)}(\bPsi_i,\bPsi_r)  $, denoted by $\bar{\bar{g}}_{(m_\x,m_\y)}(\bPsi_i,\bPsi_r) $, i.e.,
\begin{IEEEeqnarray}{ll}\label{Eq:gfunc_cont}
	\bar{\bar{g}}_{(m_\x,m_\y)}(\bPsi_i,\bPsi_r)  
	= &\tilde{\bar{g}}\int_{x=0}^{L_\x}\int_{y=0}^{L_\y}
	\e^{\frac{\jj 2\pi A_\x(\bPsi_i,\bPsi_r)}{\lambda}x}
		\e^{\frac{\jj 2\pi A_\y(\bPsi_i,\bPsi_r)}{\lambda}y}\nonumber\\
	 &\times \e^{-\frac{\jj 2\pi}{\lambda} \left[\frac{\bar{\beta}_\x}{2M_\x L_\x }x^2+\beta_{\x,m_\x} x\right]} 
	\e^{-\frac{\jj 2\pi}{\lambda} \left[\frac{\bar{\beta}_\y }{2M_\y L_\y}y^2+\beta_{\y,m_\y} y\right]}\mathrm{d}x\mathrm{d}y, \quad\,\,
\end{IEEEeqnarray}
where $ \tilde{\bar{g}}=\frac{4\pi}{\lambda^2}$ and $L_t=Q_t d_t,\,\,t\in\{\x,\y\}$. The solution to the above integrals is given in the following proposition.

\begin{prop}\label{Prop:gQuad}
	Given the proposed quadratic phase-shift design in \eqref{Eq:Quadratic} and a wave impinging on the IRS from AoA $\bPsi_i$,  the IRS response function along  AoD~$\bPsi_r$ is given by\vspace{-0.1cm}
	\begin{IEEEeqnarray}{ll}\label{Eq:gQuad}
		\bar{\bar{g}}_{(m_\x,m_\y)}(\bPsi_i,\bPsi_r) 
		= &\frac{-\jj\pi\tilde{\bar{g}}}{4\sqrt{\vartheta_\x\vartheta_\y}}
		\e^{-\jj\frac{\nu_\x^2}{4\vartheta_\x}} 
		\e^{-\jj\frac{\nu_\y^2}{4\vartheta_\y}} 
		\nonumber\\
		&\times\left[\mathrm{erfi}\left(\sqrt{\frac{\jj}{4\vartheta_\x}}\,\upsilon_\x\right)
		- \mathrm{erfi}\left(\sqrt{\frac{\jj}{4\vartheta_\x}} \,\nu_\x\right)\right]\qquad
		\nonumber\\
		&\times\left[\mathrm{erfi}\left(\sqrt{\frac{\jj}{4\vartheta_\y}}\,\upsilon_\y\right)
		- \mathrm{erfi}\left(\sqrt{\frac{\jj}{4\vartheta_\y}} \,\nu_\y\right)\right],\qquad
	\end{IEEEeqnarray}
	where $\vartheta_t=-\frac{\pi\bar{\beta}_t}{M_tL_t\lambda}$, $\nu_t=\frac{2\pi(A_t(\bPsi_i,\bPsi_r)-\beta_{t,m_t})}{\lambda}$, $\upsilon_t=\frac{2\pi(A_t(\bPsi_i,\bPsi_r)-\beta_{t,m_t}-\frac{\bar{\beta}_t}{M_t})}{\lambda}$, $t\in\{\x,\y\}$, and $\mathrm{erfi}(\cdot)$ is the imaginary error function.
\end{prop}
\begin{IEEEproof}
	The proof follows from the following integral identity \cite{wolfram2020Intquad}:
	\begin{IEEEeqnarray}{ll}\label{Eq:Integral}
		\int \e^{\jj\left(\vartheta x^2+\nu x\right)}\mathrm{d}x
		= \sqrt{\frac{\pi}{\jj 4\vartheta}} \e^{-\jj\frac{\nu^2}{4\vartheta}} 
		\mathrm{erfi}\left(\sqrt{\frac{\jj}{4\vartheta}}\left(2\vartheta x+\nu\right)\right)\quad
	\end{IEEEeqnarray}
	and the simplification of the result. 
\end{IEEEproof}

We will show in Section~\ref{Sec:sim} that the analytical expression in \eqref{Eq:gQuad} yields an accurate approximation of the IRS response function in \eqref{Eq:gfunc} particularity around the main lobe. Although  \eqref{Eq:gQuad} explicitly characterizes the dependence of the IRS response function on the codebook size $M$, substituting \eqref{Eq:gQuad} into \eqref{Eq:pEE} does not admit further simplification for the power efficiency. Therefore, we study the tradeoff between power efficiency and codebook size via simulations in Section~\ref{Sec:sim}.

\subsection{Ideal (Unattainable) Tradeoff}\label{Sec:Ideal} 

Next, we derive a simple tradeoff between power efficiency and codebook size for an idealistic scenario. We consider  the reflected beam pattern of the linear phase-shift design as the reference beam.  First, recall that $g_m(\boldsymbol{\Psi}_i,\boldsymbol{\Psi}_r)$ cannot exceed $g_{\max} = \bar{g}Q_\x Q_\y$, which is attainable by the reference beam only at its peak. Next, consider an ideal reflected beam, referred to as beam $I_1$, which has a flat main lobe with gain $g_{\max}$ covering $S=\frac{4\sin(\theta_r)}{Q_\x Q_\y}$ steradian (i.e., the maximum beamwidth for the main lobe of the reference beam along elevation angle $\theta_r$ and for all azimuth angles when $Q_\x$ and $Q_\y$ are large). The power radiated in this area is proportional to $g^2_{\max}S=4Q_\x Q_\y\sin(\theta_r)$. Now, let us construct an ideal beam $I_2$ by dividing the entire space over the IRS into $M_\x M_\y$ partitions, where $M_\x\ll Q_\x$  and $M_\y\ll Q_\y$ hold, and each subspace ($\propto\frac{2\pi}{M_\x M_\y}$ steradian) is supposed to be covered by one reflected beam. Following the law of conservation of energy, since beam $I_2$ covers an area $\frac{Q_\x Q_\y}{M_\x M_\y}$ times larger than beam $I_1$, its corresponding IRS response function is $\frac{\sqrt{M_\x M_\y}}{\sqrt{Q_\x Q_\y}}$~smaller which implies $g_m(\boldsymbol{\Psi}_i,\boldsymbol{\Psi}_r)\leq\bar{g}\sqrt{M_\x M_\y Q_\x Q_\y}$ and $\gamma(\boldsymbol{\Psi}_i,\boldsymbol{\Psi}_r)\leq\frac{M_\x M_\y} {Q_\x Q_\y}$. This simple argument provides a basic intuition regarding the tradeoff between power efficiency and codebook size.

\section{Simulation Results}\label{Sec:sim}

\begin{figure} 
	\centering\hspace{-2cm}
		\begin{minipage}{1\linewidth}
		\includegraphics[width=1.1\linewidth]{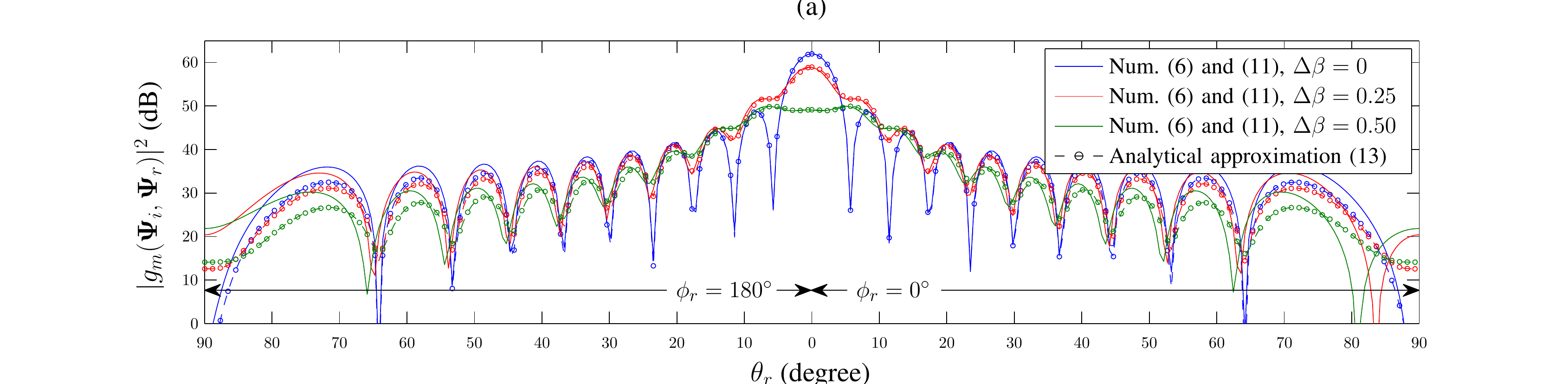}
		\vspace{-0.3cm}
	\end{minipage}
	\begin{minipage}{1\linewidth}
	\hspace{-1cm}	\includegraphics[width=1.1\linewidth]{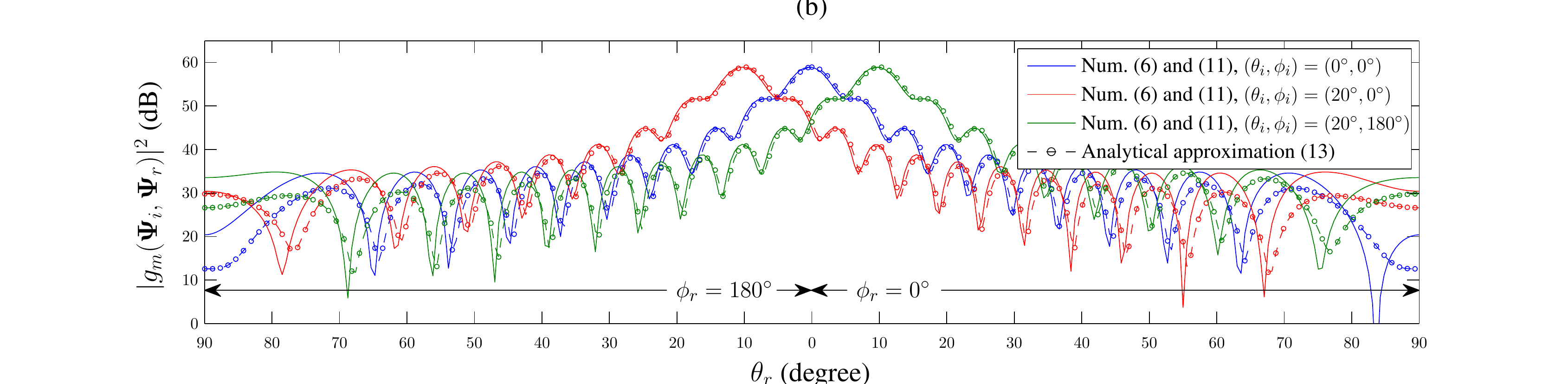}\vspace{0.1cm}
	\end{minipage}
	\begin{minipage}{1\linewidth}
	\hspace{-1cm}	\includegraphics[width=1.1\linewidth]{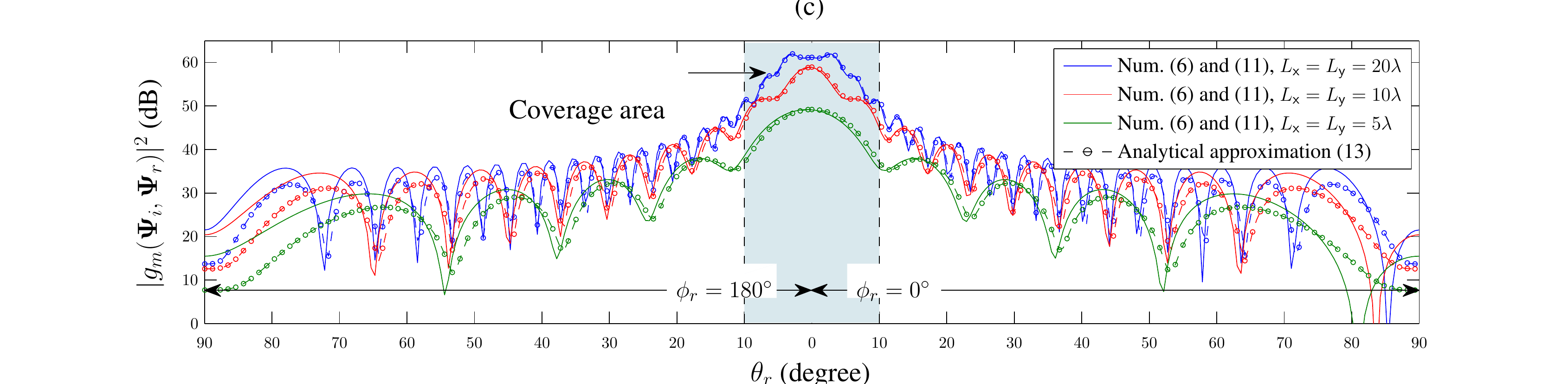}
\end{minipage}
	\vspace{-0.2cm}
	\caption{$|g_m(\boldsymbol{\Psi}_i,\boldsymbol{\Psi}_r)|^2$ in dB vs. $\theta_r$  for  $d_\x=d_\y=\frac{\lambda}{2}$. a) $(\theta_i,\phi_i)=(0,0)$,  $\phi_r\in\{0,\pi\}$, $L_\x=L_\y=10\lambda$, and different $\Delta \beta$, b) $\phi_r\in\{0,\pi\}$, $L_\x=L_\y=10\lambda$,  $\Delta \beta=0.25$, and different $(\theta_i,\phi_i)$,   c) $(\theta_i,\phi_i)=(0,0)$,  $\phi_r\in\{0,\pi\}$, $\Delta \beta=0.25$, and different $L_\x,L_\y$. \vspace{-0.5cm}} 
	\label{Fig:Quad}
\end{figure}


 In Fig.~\ref{Fig:Quad}, we plot $|g_m(\boldsymbol{\Psi}_i,\boldsymbol{\Psi}_r)|^2$ in dB vs. $\theta_r$ for a) different $\Delta\beta_{t,m}=\Delta\beta=0,025,0.5,0.75$, b)  different incident wave angles $(\theta_i,\phi_i)=(0,0),(20^\circ,0),(20^\circ,180^\circ)$, and c) different IRS size $L_\x=L_\y=5\lambda,10\lambda,20\lambda$.  Fig.~\ref{Fig:Quad}a) shows that the reflected beam becomes wider as $\Delta\beta$ increases which implies that smaller numbers of beams (i.e., smaller phase-shift codebooks) are needed to cover the entire space. On the other hand, as the beam becomes wider, the peak of the main lobe decreases which illustrates the  fundamental tradeoff between power efficiency and codebook size. Note that, in practice, not only the desired reflection direction is a priori unknown but also the angle of the incident wave. Fig.~\ref{Fig:Quad}b) shows the IRS response function  for different incident angles for a fixed phase-shift configuration. This figure suggests that for the proposed phase-shift design, a change of the incident angle translates to a change of the location of the center of the reflected beam while the shape of the reflected beam is preserved. 
 Furthermore, Fig.~\ref{Fig:Quad}c) shows that if $\Delta\beta$ (or equivalently the codebook size and the coverage area per codebook element) is kept fixed, increasing the IRS size simply boosts $|g_m(\boldsymbol{\Psi}_i,\boldsymbol{\Psi}_r)|$ within the coverage area of each codebook element.
 Finally, Fig.~\ref{Fig:Quad} reveals that there is a good agreement between the IRS response function computed numerically from \eqref{Eq:gfunc} and \eqref{Eq:Quadratic} and the approximated analytical expression in  \eqref{Eq:gQuad}.

\begin{figure}[!t]
	\centering\hspace{-2cm}
	\begin{minipage}{1\linewidth}
	\includegraphics[width=1.1\linewidth]{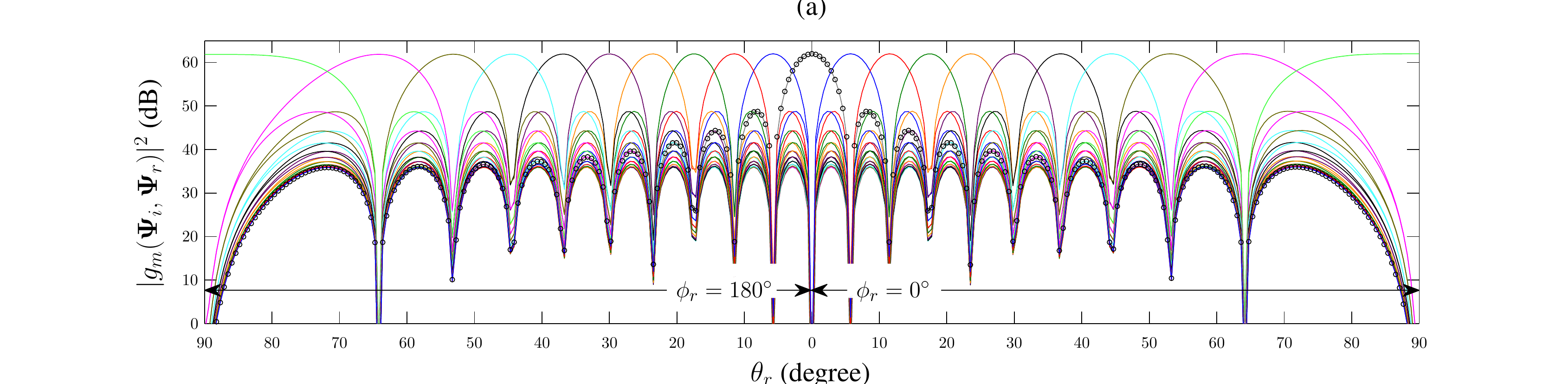}
\vspace{-0.3cm}
	\end{minipage}
\begin{minipage}{1\linewidth}
		\hspace{-1cm}
	\includegraphics[width=1.1\linewidth]{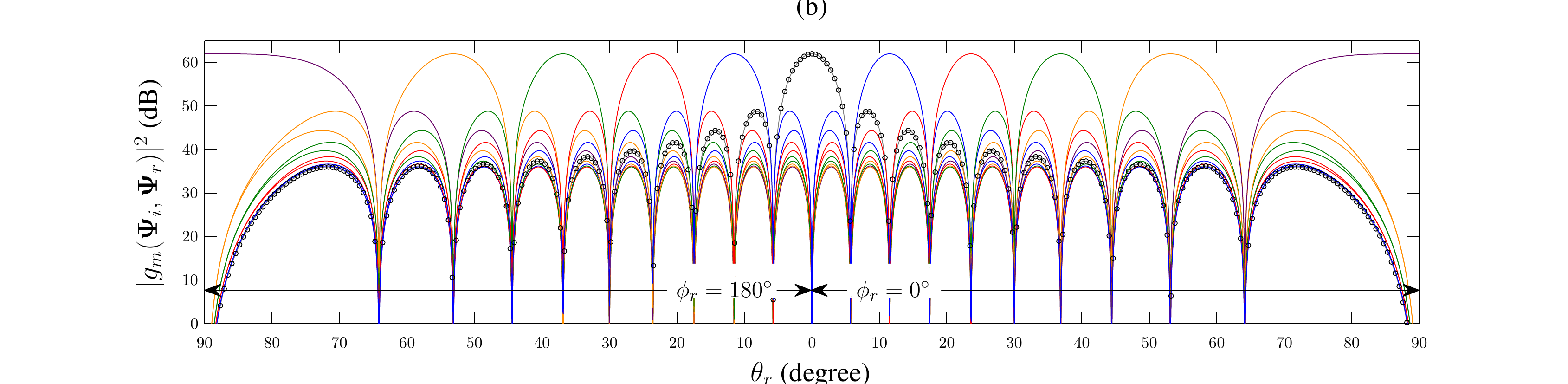}
\vspace{-0.3cm}
\end{minipage}
	\begin{minipage}{1\linewidth}
			\hspace{-1cm}
	\includegraphics[width=1.1\linewidth]{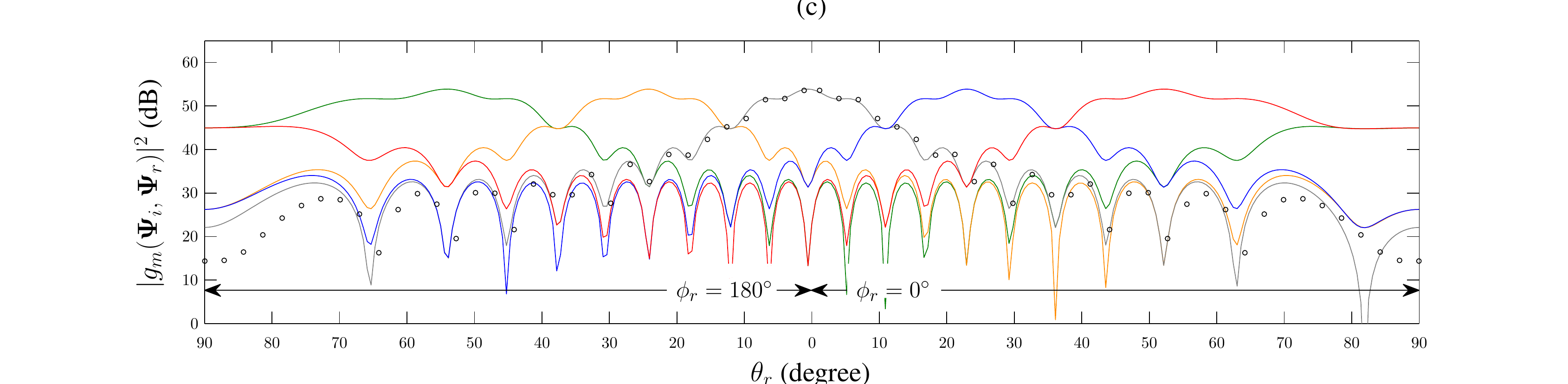}
\end{minipage}
	\vspace{-0.2cm}
	\caption{$|g_m(\boldsymbol{\Psi}_i,\boldsymbol{\Psi}_r)|^2$ in dB vs. $\theta_r$  for $(\theta_i,\phi_i)=(0,0)$,  $\phi_r=\{0,\pi\}$, $L_\x=L_\y=10\lambda$, $d_\x=d_\y=\frac{\lambda}{2}$. a) DFT-based phase-shift design $M=400$, b) linear phase-shift design in \cite{najafi2020intelligent} with $M=100$, and c) proposed quadratic phase-shift design $M=25$. The markers are analytical results from Lemma~\ref{Lem:gLinear} and Proposition~\ref{Prop:gQuad}. \vspace{-0.5cm}} 
	\label{Fig:Beam}
\end{figure}

In Fig.~\ref{Fig:Beam}, we plot $|g_m(\boldsymbol{\Psi}_i,\boldsymbol{\Psi}_r)|^2$ in dB vs. $\theta_r$ for a) the DFT-based phase-shift design in \eqref{Eq:DFT}, b) the linear phase-shift design in \eqref{Eq:Linear}, and c) the proposed quadratic phase-shift design in \eqref{Eq:Quadratic}. Since we have  $Q_\x Q_\y=400$~unit~cells, the size of the DFT-based codebook is  $M=400$, which is too large for practical implementation. For the linear design in \eqref{Eq:Linear}, $M>\frac{Q_\x Q_\y}{4}=100$ has to hold to ensure  a large $\gamma(\boldsymbol{\Psi}_i,\boldsymbol{\Psi}_r),\,\,\forall\boldsymbol{\Psi}_i,\boldsymbol{\Psi}_r$. In particular, we observe from Fig.~\ref{Fig:Beam}b) that  for $M=100$, there are AoDs without coverage, i.e., $\gamma(\boldsymbol{\Psi}_i,\boldsymbol{\Psi}_r)=0$. In contrast, the proposed design is able to cover the entire space with only $M=25$ which underlines its practical~relevance. 

\begin{figure} 
	\centering
	\includegraphics[width=1\linewidth]{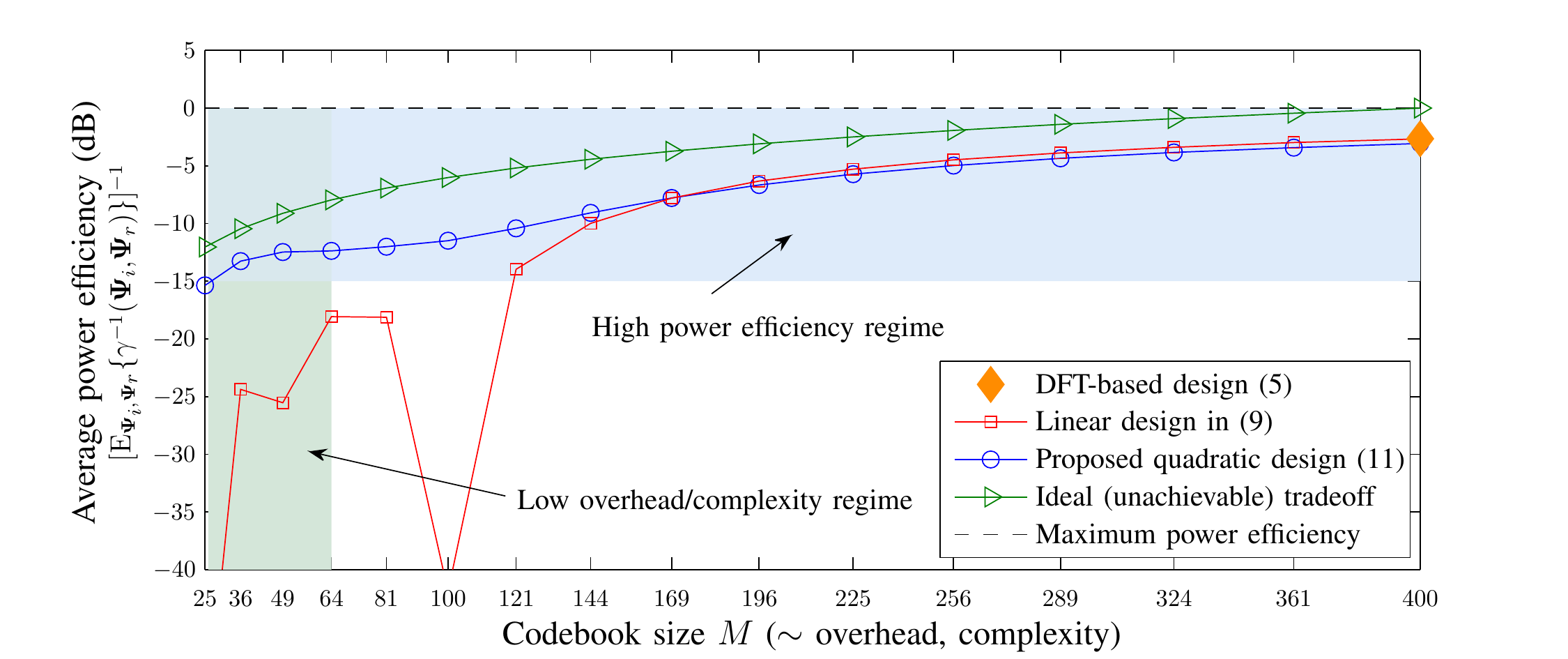}
	\vspace{-0.7cm}
	\caption{Power efficiency in dB vs. phase-shift codebook size $M$  for  $L_\x=L_\y=10\lambda$, $d_\x=d_\y=\frac{\lambda}{2}$.  \vspace{-0.5cm}} 
	\label{Fig:Tradeoff}
\end{figure}

Finally, in Fig.~\ref{Fig:Tradeoff}, we illustrate the tradeoff between power efficiency and codebook size $M$. The results in Fig.~\ref{Fig:Tradeoff} were obtained by averaging over $10^5$ uniformly random realizations of the AoAs and AoDs for the IRS. Assuming that we are interested in ensuring a given received power, the BS's transmit power has to be proportional to $\gamma^{-1}(\boldsymbol{\Psi}_i,\boldsymbol{\Psi}_r)$. Therefore,  $\mathrm{E}_{\boldsymbol{\Psi}_i,\boldsymbol{\Psi}_r}\{\gamma^{-1}(\boldsymbol{\Psi}_i,\boldsymbol{\Psi}_r)\}$ can be seen as a measure for the average  BS transmit power, where $\mathrm{E}\{\cdot\}$ denotes  expectation. In Fig.~\ref{Fig:Tradeoff}, we show  $[\mathrm{E}_{\boldsymbol{\Psi}_i,\boldsymbol{\Psi}_r}\{\gamma^{-1}(\boldsymbol{\Psi}_i,\boldsymbol{\Psi}_r)\}]^{-1}$ vs. the codebook size which is a measure for the channel estimation overhead and online optimization complexity. As can be seen from this figure, the DFT-based design offers a good power efficiency at the cost of a large codebook size. The linear design in \cite{najafi2020intelligent} offers a tunable tradeoff between power efficiency and codebook size; however, the power efficiency becomes very small for small codebook sizes which is the regime of interest for many practical applications.  In fact, for $M\leq100$, there exist AoAs and AoDs which are  supported only by the side lobes of $\gamma(\boldsymbol{\Psi}_i,\boldsymbol{\Psi}_r),\,\,\forall m$, cf. Corollary~\ref{Corol:Pefflinear},  which is the reason for the fluctuation of the corresponding curve in Fig.~\ref{Fig:Tradeoff}. On the other hand, the proposed quadratic phase-shift design offers a much better tradeoff as it  enables a high power efficiency even for very small codebook sizes. Furthermore, the tradeoff obtained with the proposed phase-shift design is close to the unattainable tradeoff of the ideal scenario discussed in~Section~\ref{Sec:Ideal}.

\vspace{-0.3cm}
\bibliographystyle{IEEEtran}
\bibliography{References}
\end{document}